\documentclass[%
 reprint,
 showpacs,
 amsmath,amssymb,
 aps,
]{revtex4-1}
\usepackage{amsmath}
\usepackage{cases}
\usepackage{amssymb}
\usepackage{CJK}% Chinese character support
\usepackage{graphicx}% Include figure files
\usepackage{dcolumn}% Align table columns on decimal point
\usepackage{bm}% bold math
\usepackage{color}
\usepackage[pagewise]{lineno}% add line numbers

\begin{document}
\begin{CJK*}{GBK}{} % Use default fonts from CJK (see below)

\preprint{APS/123-QED}

\title{An Empirical Scaling formula for fragments production in projectile fragmentation reactions
}% Force line breaks with \\
%\thanks{A footnote to the article title}%
\author{Yi-Dan SONG}%$^{1}$}
\author{Chun-Wang MA}%$^{1}$}
\thanks{Email address: machunwang@126.com}
\author{Xue-Ying LIU}
\author{Hui-Ling WEI}
%\homepage{http://www.researcherid.com/rid/B-3848-2009}
\affiliation{%$^{1}$
Institute of Particle and Nuclear Physics, Henan Normal University, \textit{Xinxiang 453007}, China
}

% \altaffiliation[Also at ]{Physics Department, XYZ University.}%Lines break automatically or can be forced with \\
%\author{Second Author}%
% \email{Second.Author@institution.edu}

%\author{Charlie Author}
% \homepage{http://www.Second.institution.edu/~Charlie.Author}
%\affiliation{
% Second institution and/or address\\
% This line break forced% with \\
%}%
%\affiliation{
% Third institution, the second for Charlie Author
%}%

\date{\today}
%Start preparation 6th, Oct. 2016., First draft 9th, Oct. 2016.

%\linenumbers% show line numbers
\begin{abstract}
A scaling phenomenon in the cross section for fragments has been found in the projectile fragmentation reaction, and an empirical scaling formula is proposed by considering the dependence of cross section on the size and asymmetry of the reaction system and the fragment itself. Furthermore, the empirical scaling formula is used to predict the production of fragment in the $^{68}$Ni/$^{69}$Cu/$^{72}$Zn + $^9$Be reactions around 90$A$ MeV. Compared to the results calculated by the statistical abrasion ablation model and the {\sc epax3} parameterizations, the empirical scaling formula can better reproduce the measured fragments. The empirical scaling formula can be used to predict the yield for fragment in projectile fragmentation reaction.
\end{abstract}

% PACS, the Physics and Astronomy
                             % Classification Scheme.
                             % 25.70.Pq Multifragment emission and correlations
                             % 24.60.-k Statistical theory and fluctuations
                             % 25.70.Mn Projectile and target fragmentation
                             % 21.65.Cd Asymmetric matter, neutron matter
%Use showkeys class option if keyword
                              %display desired
\maketitle
\end{CJK*}

%\tableofcontents
%\linenumbers% show line numbers

\section{introduction}
Projectile fragmentation is one kind of violent nuclear collisions by bombarding an accelerated nucleus on a target nucleus, in which a variety of fragments from the proton-rich side to the neutron-rich side can be produced. Many models have been developed to predict the yield for fragments in projectile fragmentation reactions. For examples, the parameterizations {\sc epax} \cite{EPAX} and the improved versions {\sc epax2} and {\sc epax3} \cite{EPAX2,EPAX3,Mocko06,ZXH12EPAX}, the statistical models  \cite{SAA94,SAA91,SAA00F,SMM00PRC,SMM01CPC,MaCW09PRC}, the thermal dynamics models \cite{MFM82,MFM84,GCE01,GCE07}, the macroscopic-microscopic heavy ion phase space exploration ({\sc HIPSE}) model \cite{hipse,Mocko06,Mocko08}, and the more complex transport models \cite{QMD91,QMD00,Ono99AMD,Mocko08,Huang10IYR,Ma15CPL,IBD15AMD}. Some scaling phenomena for fragments produced in projectile fragmentation reactions have been found, for example, the isoscaling phenomenon \cite{Tsang01Iso}, the $m$-scaling phenomena ($m = \frac{N-Z}{A}$ of a fragment) \cite{H-mscaling}, and the isobaric ratio difference scaling phenomenon \cite{info15PLB,info16JPG}. These scaling phenomena for fragments reflect general properties of the reaction, which are used as probes to study the nuclear matter in heavy-ion collisions \cite{ppnp-sym,BARep}.

The success of the {\sc epax2} and {\sc epax3} parameterizations indicates that there are general rules to describe the yield for fragments in projectile fragmentation reaction, while the complex parameters make it not easy to understand the physics in them. It is useful to find a simple description for fragment production in projectile fragmentation reaction. The cross section for fragments in the 140$A$ MeV $^{40, 48}$Ca + $^{9}$Be/$^{181}$Ta and $^{58, 64}$Ni + $^{9}$Be/$^{181}$Ta reactions have been measured by M. Mocko \textit{et al} at the National Superconducting Cyclotron Laboratory (NSCL) in Michigan State University \cite{Mocko06}. The high quality data provide an opportunity to study the general law of fragment production in the projectile fragmentation reaction. In this work, an empirical scaling formula for fragments will be firstly discussed by studying the measured fragments in the 140$A$ MeV $^{40, 48}$Ca + $^{9}$Be and $^{58, 64}$Ni + $^{9}$Be reactions, and then the empirical scaling formula is used to predict the cross section for fragments in the measured $^{68}$Ni/$^{69}$Cu/$^{72}$Zn + $^9$Be reactions around 90$A$ MeV, which were also measured at NSCL by S. Lukyanov \textit{et al} \cite{Luk09}.

The letter is organized as follows. In Sec. \ref{method} the formulism for the empirical scaling formula is proposed combining the general physical ideas of fragment production. In Sec. \ref{RAD}, with the $^{40, 48}$Ca + $^{9}$Be and $^{58, 64}$Ni + $^{9}$Be reactions being analyzed, the parameters in the empirical scaling formula is discussed. The empirical scaling formula is further used to predict the cross section for fragments in the $^{68}$Ni/$^{69}$Cu/$^{72}$Zn + $^9$Be reactions. A brief summary is presented in Sec. \ref{summary}.

\section{formulism}
\label{method}
\begin{figure*}[htbp]
%\centering
\includegraphics%[width=\columnwidth]
[width=16.8cm]{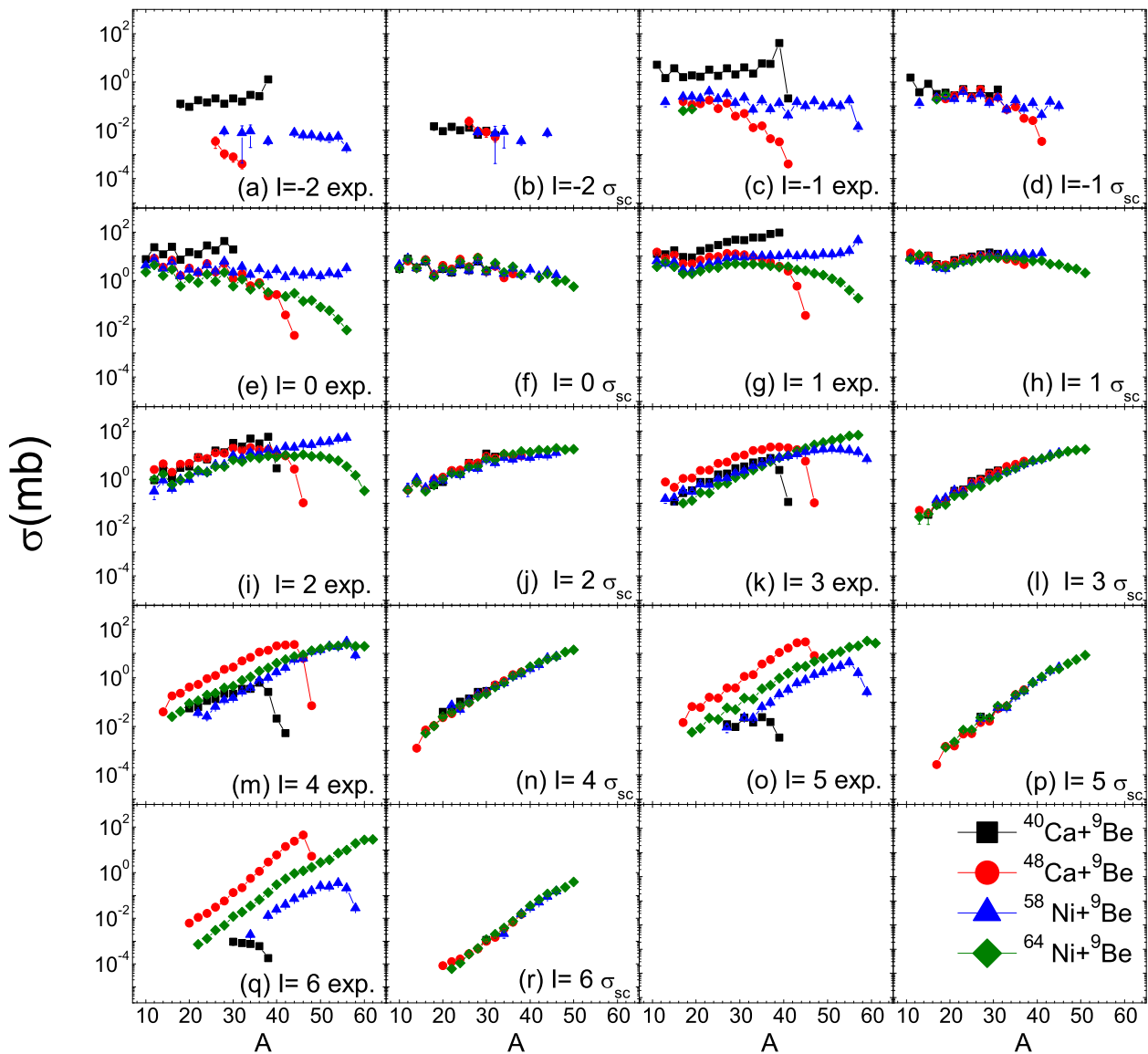}
\caption{\label{CaNiBe} (Color online) The measured and scaled cross section for fragments produced in the 140$A$ MeV $^{40, 48}$Ca + $^{9}$Be and $^{58, 64}$Ni + $^{9}$Be reactions. Only the results for the fragments with $A_f/A_P \leq 0.8$ are plotted. The measured results are taken from \cite{Mocko06}.
}
\end{figure*}

To describe the formulism, some notations are defined. The projectile nucleus, the fragment nucleus, the yield for fragment will be used. Z$_{i}$, N$_{i}$, and A$_{i}$ are used to denote the the charge, neutron and mass numbers, in which $i= R, P$ and refer to the projectile nuclei of the reference and studied system, and $i = f$ refers to fragment in the studied system, respectively.
It is known that the fragment production in the projectile fragmentation reaction depends on the size and asymmetry of the system, as well as the incident energy of the reaction. At the same time, the asymmetry of fragment itself also influences its production. The isospin effect and its disappearance has been discussed by dividing the projectile nucleus into the core part and the surface part, which correspond to the central collisions and the peripheral collisions, respectively \cite{MaCW09PRC}. The similarity of yields for fragments with relatively small mass numbers denotes that in central collisions the neutron and proton densities are similar, while the large difference between the densities of neutron and proton in the surface region of projectile nucleus accounts for the isospin effect \cite{MaCW09PRC}. This is further illustrated by the isobaric scaling phenomenon between neutron-rich fragments \cite{info15PLB,info16JPG}. The neutron skin of the very neutron-rich projectile nucleus has a significant influence on the yield of fragment which has a mass number very close to that of the projectile, i.e., produced in the peripheral collisions \cite{MaCW09PRC,MACW10JPG,MaCW10PRC,IBD14Ca,Ma13finite}. To avoid the influence of neutron skin structure, only the fragments with $A_{f}/A_{P} \leq$ 0.8 are analyzed in this work. According to the neutron excess ($I \equiv N - Z$), the distribution for fragments from $I =$ -2 to 6 are plotted in Fig. \ref{CaNiBe}.

From Fig. \ref{CaNiBe}, it is clearly shown that the measured cross section for fragment depends on the mass and asymmetry of the projectile nucleus, as well as the mass and $I$ of the fragment itself. For the fragments with $I \leq$ 1, the cross section depends less on $A_f$ except in the very neutron-rich $^{48}$Ca induced reaction. While the mass dependence of cross section becomes more obvious as $I$ increases. The proton-rich and very neutron-rich fragments show an very obvious size dependence of the system. To find the scaling phenomena of the cross section for fragments in difference reactions, the main tasks are to eliminate the dependence of the yield fragment on the size and isospin of the system, as well as the dependence of mass and isospin of the fragment itself. Considering the above effects, the following empirical formula is proposed to calculate the scaled cross section ($\sigma_{sc}$) from the measured one ($\sigma_{exp}$),
\begin{equation}\label{mainFor}
\sigma_{sc}=\sigma_{exp}\cdot(\frac{A_{P}}{A_{R}})
\cdot(\frac{N_{P}}{N_{R}}\frac{Z_{P}}{Z_{R}})
\cdot(\frac{N_{P}}{Z_{P}}\frac{N_{f}}{Z_{f}})^{a}
\cdot(\frac{A_{r}}{A_{f}})^{b}\cdot K,
\end{equation}
in which the second term $\frac{A_{P}}{A_{R}}$ deals with the size dependence of the systems. The term $\frac{N_{P}}{N_{R}}\frac{Z_{P}}{Z_{P}}$ deals with the isospin effect, which is induced by the difference of asymmetry between the studied and the reference systems. The term $(\frac{N_{P}}{Z_{P}}\frac{N_{f}}{Z_{f}})^{a}$ deals with the difference of asymmetry between the projectile and the fragment nuclei in the studied reaction. A reference fragment ($A_{r}$) is also adopted to deal with the size effect between fragments. The term $(A_{r}/A_{f})^{b}$ is related to the isobaric effect of fragment, which is influenced by the neutron excess of fragment. The parameters $a$, $b$ and $K$ are adjusted to improve the scaling phenomenon, which will be explained combining the analysis of experimental data in Sec. \ref{RAD}.

\section{results and discussion}
\label{RAD}
In this section, the values for $a$, $b$ and $K$ will be discussed. $(\frac{N_{P}}{Z_{P}}\frac{N_{f}}{Z_{f}})^{a}$ is used to adjust the isospin between the projectile and the fragment. The values for $a$ is influenced by $I$ of fragment,
\begin{equation}\label{avaluep}
a=\left\{
\begin{aligned}
 & ~~~1,~~ \hspace{1.cm} I \leq 1, \\
 & -1, \hspace{1.2cm}  1 < I \leq 5, \\
 & ~~~1,~~ \hspace{1.cm} I > 5,
\end{aligned}
\right.
\end{equation}
$a =$ 1 and -1 denote that the fragment is difficult or easy to be produced, respectively, which reflects the fact that the cross section of fragment decreases fast as the asymmetry of fragment increases. In this work, the $^{58}$Ni + $^{9}$Be reaction, which is a symmetric system, is taken as the reference system in Eq. (\ref{mainFor}).

The fragments have been found to well obey an $m-$scaling ($m = I/A$) \cite{H-mscaling}. The $I = 0$ fragments are well normalized to the yield of $^{12}$C in different reactions. The $m-$ scaling phenomenon means that the cross section for fragment depends not only on $I$ but also on $A$ of fragment. $(\frac{A_{r}}{A_{f}})^{b}$ is introduced in Eq. (\ref{mainFor}), for which $(\frac{A_{r}}{A_{f}})$ reflects the size effect between the fragments, and $b$ is used to reflect the strength as well as the $m-$scaling of fragment. For a fragment with $I \leq$ 1,
\begin{equation}\label{bvaluep}
b=\left\{
\begin{aligned}
 & |I|, \hspace{0.6cm} (N/Z)_{P} < (N/Z)_{R}, \\
 & 0,   \hspace{0.8cm} (N/Z)_{R},\\
 & I - 1, \hspace{0.2cm} (N/Z)_{P} > (N/Z)_{R}.
\end{aligned}
\right.
\end{equation}
For a neutron-rich fragment with 1 $< I \leq$ 5,
\begin{equation}\label{bvaluen1}
b = -1.
\end{equation}
For a neutron-rich fragment with $I >$ 5,
\begin{equation}\label{bvaluen2}
b = 1.
\end{equation}

$K$ is further introduced to modify the effect of neutron excess. For a fragment with $I\leq$ 1,
\begin{equation}\label{Kvaluen1}
K = 1,
\end{equation}
which is independent of $I$ and the asymmetry of the projectile nucleus.
For a fragment with $1 < I \leq 5$, $K$ depends both on $I$ and the asymmetry of the projectile nucleus,
\begin{equation}\label{Kvalue2}
K=\left\{
\begin{aligned}
 & 2^{(I - 3)},\hspace{0.6cm} (N/Z)_{P} < (N/Z)_{R}, \\
 & I,          \hspace{1.3cm}   (N/Z)_{R},\\
 & 1/2^{(I - 1)}   \hspace{0.4cm} (N/Z)_{P} > (N/Z)_{R}.
\end{aligned}
\right.
\end{equation}
For the more neutron-rich fragment of $I >$ 5,
\begin{equation}\label{Kvalue3}
K=\left\{
\begin{aligned}
 & 1,     \hspace{1.2cm} (N/Z)_{P} < (N/Z)_{R},\\
 & 1/4,   \hspace{0.9cm}  (N/Z)_{R},\\
 & 1/2^{6},\hspace{0.8cm} (N/Z)_{P} > (N/Z)_{R}.
\end{aligned}
\right.
\end{equation}
In a neutron-rich system, the production for a neutron-rich fragment is enhanced. $K$ is introduced to reduce this enhancement for the neutron-rich fragment, which makes $K$ for the neutron-rich fragment become smaller for the neutron-rich system.
For the symmetric fragments and proton-rich fragments, i.e., $I \leq$ 1, the values of $a, b$ and $K$ are uniform. While for the neutron-rich fragments, the values for $K$ are relatively complex.

\begin{figure*}[htbp]
%\centering
\includegraphics%[width=\columnwidth]
[width=16.8cm]{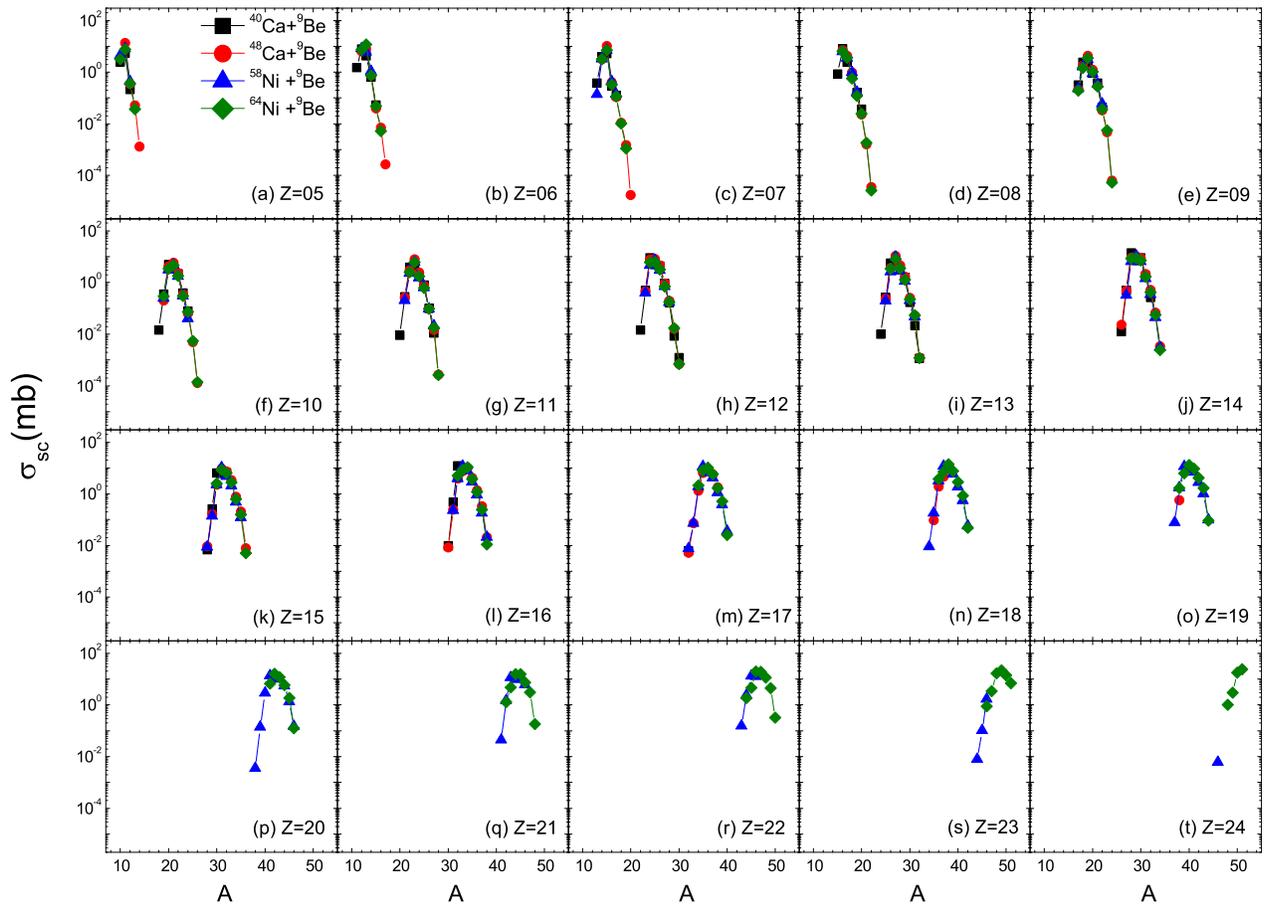}
\caption{\label{ScaledIstp} (Color online) The isotopic distribution of the scaled cross section for fragments produced in the 140$A$ MeV $^{40, 48}$Ca + $^{9}$Be and $^{58, 64}$Ni + $^{9}$Be reactions.
}
\end{figure*}

The measured cross section for fragments produced in the $^{40, 48}$Ca + $^{9}$Be and $^{58, 64}$Ni + $^{9}$Be reactions have been calculated using Eq. (\ref{mainFor}), and are also plotted in Fig. \ref{CaNiBe}. It can be seen that the calculated yield for fragments are well scaled for each $I$. Usually, the isotopic distribution for fragment yield is used to study the isospin effect in projectile fragmentation reactions \cite{MaCW09PRC,MACW10JPG}. The similar isotopic distribution in different reactions reflects a similar neutron and proton density distributions between the systems. The isotopic distributions for the scaled cross sections are plotted in Fig. \ref{ScaledIstp}. From $Z =5$ to $Z = 24$, it is seen that the isotopic distribution for the $^{40, 48}$Ca + $^9$Be and $^{58, 64}$Ni + $^9$Be reaction overlap after the scaling calculation. The isospin effect in the experimental isotopic distributions \cite{MaCW09PRC} has been cancelled out, which indicates that the reaction systems with different asymmetries can be scaled.

The values of $N/Z$ for $^{40}$Ca, $^{58, 64}$Ni and $^{48}$Ca are 1.0, 1.071, 1.214, and 1.4, respectively, which cover a relatively wide range of asymmetry. The scaling phenomenon for fragments with different $I$ suggests that the fragment production in similar reaction can be predicted. The cross section for fragments in the 93.7$A$ MeV $^{64}$Ni + $^{9}$Be, 94.3$A$ MeV $^{68}$Ni + $^{9}$Be, 98.1$A$ MeV $^{69}$Cu + $^{9}$Be and 95.4$A$ MeV $^{72}$Zn + $^{9}$Be reactions \cite{Luk09}. The comparison between the measured cross section for fragments in the $^{64}$Ni + $^{9}$Be at 140$A$ MeV and 93.7$A$ MeV reaction show that, except for the proton-rich or very neutron-rich fragments, the isotopic distributions are very similar \cite{Luk09}. It means that the incident energy of the reaction has very little influence on the fragment production at the two incident energies. The cross section for fragments in the $^{68}$Ni/$^{69}$Cu/$^{72}$Zn + $^{9}$Be reactions are predicted using the empirical scaling formula. Meanwhile, the cross section for fragments are also calculated using a modified statistical abrasion ablation (SAA) model \cite{SAA94,SAA91,SAA00F,MaCW09PRC} and the {\sc epax3} parameterizations \cite{EPAX3}. In the SAA model calculation, the free space nucleus-nucleus cross sections and the Fermi-type density distribution are adopted, which can well reproduce the fragment productions in projectile fragmentation reactions \cite{MaCW09PRC,Ma13finite,Ma15CPL}.

The $N/Z$ of $^{69}$Cu, $^{72}$Zn and $^{68}$Ni are 1.379, 1.40, and 1.429, respectively, for which $(N/Z)_P > (N/Z)_R$. For the $^{69}$Cu, $^{72}$Zn and $^{68}$Ni reactions, the same values for $a$ and $b$ are used. $K$ has been changed for the $I \leq$ 1 and $I >$ 5 fragments by considering the relative large difference between the proton-rich and very neutron-rich fragments for the $^{64}$Ni reactions at 90 and 140$A$ MeV, which are,
\begin{equation}\label{KNiCuZn}
K = \left\{
\begin{aligned}
 & 4,     \hspace{1.5cm} I \leq 1,\\
 & 1/2^{(I - 2)},   \hspace{0.4cm} I \geq 5,
\end{aligned}
\right.
\end{equation}

\begin{figure*}[htbp]
%\centering
\includegraphics%[width=\columnwidth]
[width=16.8cm]{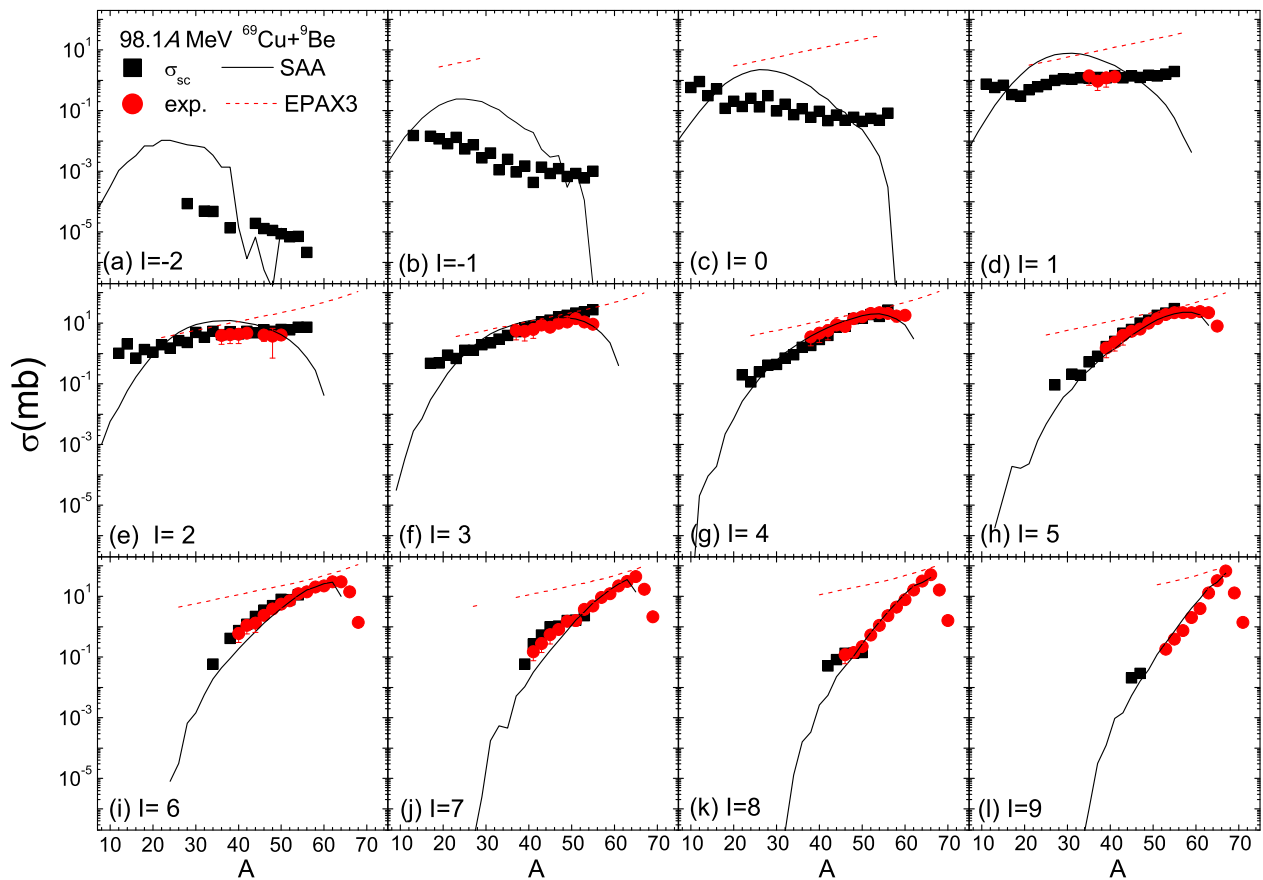}
\caption{\label{Cu69} (Color online) The cross section for fragments produced in the 94.3$A$ MeV $^{68}$Ni + $^{9}$Be, 98.1$A$ MeV $^{69}$Cu + $^{9}$Be and 95.4$A$ MeV $^{72}$Zn + $^{9}$Be reactions. The circles and squares denote the measured cross section for fragments \cite{Luk09} and the predicted results using Eq. (\ref{mainFor}). The solid and dashed lines denote the calculated cross section using the modified statistical abrasion ablation model and the {\sc epax3} parameterizations.
}
\end{figure*}

The measured cross section for fragments in the 98.1$A$ MeV $^{69}$Cu + $^{9}$Be reactions, as well as the predicted results by Eq. (\ref{mainFor}), the SAA model and the {\sc epax3} parameterizations are compared in Fig. \ref{Cu69}. It can be seen that for most of the fragments, the {\sc epax3} parameterizations overestimate the measured results. For fragments of $I \geq 3$, the SAA predictions agree with the measured results. No measured fragments with $I \leq$ 0 have been found in the experiments. The {\sc epax3} parameterizations predict a larger cross section than the results of the SAA model and Eq. (\ref{mainFor}). The predicted results by Eq. (\ref{mainFor}) agree with the measured results very well except some of the fragments. For most of the fragments with $I \geq 4$, the predicted results by Eq. (\ref{mainFor}) and the SAA model agree well. While for the fragments with $I < 4$, the trend of the distribution for fragments between the predicted results for Eq. (\ref{mainFor}) and SAA are different despite the fact that some of them are similar.

\begin{figure*}[htbp]
%\centering
\includegraphics%[width=\columnwidth]
[width=16.cm]{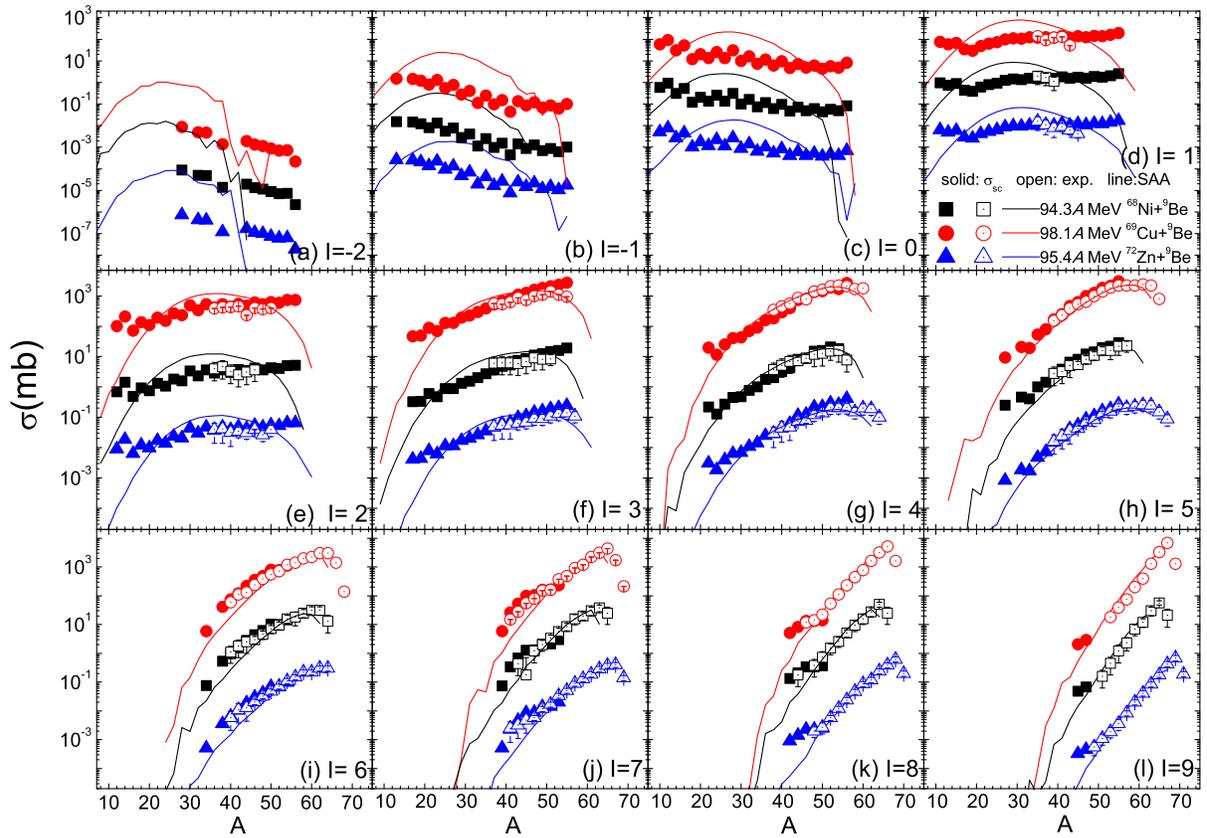}
\caption{\label{ComNiZnCu} (Color online) The isotopic distributions for the fragments produced in the 94.3$A$ MeV $^{68}$Ni + $^{9}$Be, 98.1$A$ MeV $^{69}$Cu + $^{9}$Be and 95.4$A$ MeV $^{72}$Zn + $^{9}$Be reactions. The circles and squares denote the measured cross section for fragments \cite{Luk09} and the predicted results using Eq. (\ref{mainFor}). The results for the $^{69}$Cu reaction are multiplied by 100, and those for the $^{72}$Zn reaction are divided by 100.
}
\end{figure*}

The cross sections for fragments produced in the 94.3$A$ MeV $^{68}$Ni + $^{9}$Be and 95.4$A$ MeV $^{72}$Zn + $^{9}$Be reactions have also been calculated by the SAA model and the {\sc epax3} parameterizations. The results are similar to those for the 98.1$A$ MeV $^{69}$Cu + $^{9}$Be reaction. For simplification, the predicted results by Eq. (\ref{mainFor}), the SAA model for the measured fragment in the 94.3$A$ MeV $^{68}$Ni + $^{9}$Be, 98.1$A$ MeV $^{69}$Cu + $^{9}$Be and 95.4$A$ MeV $^{72}$Zn + $^{9}$Be are compared in Fig. \ref{ComNiZnCu}. For the $I \leq$ 0 fragments, the predicted cross section by Eq. (\ref{mainFor}) decrease slightly as $A$ of fragment increases. For most of the fragments, the predicted results by Eq. (\ref{mainFor}) agree well with the measured fragments. In general, the SAA model pronounces larger cross sections than the measured results and those by the Eq. (\ref{mainFor}) for the $I \leq$ 2 fragments, while it reproduces the measured results well for the $I >$ 3 fragments. It can be concluded that the empirical scaling formula of Eq. (\ref{mainFor}) can well reproduce the cross section for fragments around the incident energy of 100$A$ MeV, which is better than the SAA model and the {\sc epax} parameterizations.

\section{summary}
\label{summary}
An empirical formula is proposed to study the scaling phenomena of fragment production in the projectile fragmentation reactions. The measured fragments in the 140$A$ MeV $^{40, 48}$Ca + $^{9}$Be and $^{58, 64}$Ni + $^{9}$Be reactions are used to obtain the scaling formula. The fragments have been limited to $A_f/A_p \leq$ 0.8 to avoid the fragments produced in the peripheral collisions. By considering the dependence of cross section on the size and asymmetry of the system, as well as the mass of fragment and asymmetry of the fragment itself, the empirical scaling formula can well scale the cross section for fragments in the $^{40, 48}$Ca and $^{58, 64}$Ni reactions. Furthermore, the empirical scaling formula is also used to predict the measured fragments in the 94.3$A$ MeV $^{68}$Ni + $^{9}$Be, 98.1$A$ MeV $^{69}$Cu + $^{9}$Be and 95.4$A$ MeV $^{72}$Zn + $^{9}$Be reactions measured at NSCL. By comparing the measured cross section for fragments, the predicted results by Eq. (\ref{mainFor}) and the SAA models, it is found that the empirical scaling formula has a good quality to predict the fragments in projectile fragmentation reactions around the incident energy of 100$A$ MeV.

\begin{acknowledgments}
This work is partially supported by the Program for Science and Technology Innovation Talents at Universities of Henan Province (13HASTIT046), the Natural and Science Foundation in Henan Province (grant No. 162300410179), and the Program for the Excellent Youth at Henan Normal University (grant No. 154100510007).
\end{acknowledgments}

\end{document}